\documentclass[sigconf, final]{acmart}
\usepackage{tikz}
\usepackage{booktabs}
\usepackage{xcolor}
\usepackage[normalem]{ulem}
\usepackage{capt-of}
\usepackage{listings}
\usepackage{tikz}
\usepackage{xspace}
\usetikzlibrary{decorations.pathmorphing, arrows.meta, positioning, calc}

\lstdefinestyle{prompt}{
  basicstyle=\ttfamily\scriptsize,
  breaklines=true,
  breakatwhitespace=false,
  frame=single,
  rulecolor=\color{gray!40},
  backgroundcolor=\color{gray!5},
  columns=fullflexible,
  keepspaces=true,
  aboveskip=6pt,
  belowskip=6pt,
  xleftmargin=4pt,
  xrightmargin=4pt,
}

\usetikzlibrary{shapes.geometric, arrows.meta, positioning, calc}

\AtBeginDocument{%
  }

\acmConference[Agent4IR' 26 KDD]{AI Agent for Information Retrieval (Agent4IR)
Co-located with KDD 2026}{August 9th-13th,  2026}{Jeju, Korea}

\begin{document}

\title{Conversational Query Engine for Mixed-Modality Heterogeneous Enterprise Data Sources}

\author{Darshita Rathore}
\orcid{0000-0003-0430-1129}
\email{drathore@paypal.com}
\affiliation{%
  \institution{PayPal Artificial Intelligence}
  \city{Bengaluru}
  \country{India}
}

\author{Vineet Kumar}
\orcid{0000-0001-8364-3981}
\email{vkumar32@paypal.com}
\affiliation{%
  \institution{PayPal Artificial Intelligence}
  \city{Bengaluru}
  \country{India}
}

\author{Vaibhav Singal}
\orcid{0009-0001-5472-2354}
\email{vasingal@paypal.com}
\affiliation{%
  \institution{PayPal Artificial Intelligence}
  \city{Bengaluru}
  \country{India}
}

\author{Ankur Vivek Singh}
\orcid{0009-0009-3904-547X}
\email{aviveksingh@paypal.com}
\affiliation{%
  \institution{PayPal Artificial Intelligence}
  \city{Bengaluru}
  \country{India}
}

\author{Anindya Moitra}
\orcid{0009-0007-8165-0003}
\email{amoitra@paypal.com}
\affiliation{%
  \institution{PayPal Artificial Intelligence}
  \city{Bengaluru}
  \country{India}
}

\begin{abstract}
Enterprise business intelligence queries span structured warehouses and unstructured document repositories -- modalities with fundamentally different access methods, cost profiles, and correctness semantics. Existing AI-enabled interfaces force users to select the right tool: NL2SQL systems cannot reason over slide decks, and RAG pipelines lack access to live warehouse tables.

We present \textbf{COGNI}, a production conversational BI system that treats natural-language analytics as a heterogeneous query processing problem, organized as four architectural layers. First, an \textbf{\emph{indexing layer}} implements slide-adaptive chunking --- recursive chunking for plain-text slides, hierarchical chunking for structured content such as tables, charts, and key-value blocks - achieving 88.3\% on our internal enterprise benchmark. Second, a \textbf{\emph{routing layer}} built on a LoRA fine-tuned Qwen-2.5-1.5B-Instruct model that produces a dual output - modality decision and complexity assessment --- at 93.8\% accuracy and approximately $7\times$ lower cost than frontier-model. Third, a \textbf{\emph{retrieval layer}} executes complexity-adaptive pipelines: a self-correcting NL2SQL agent at 93.9\% G-Eval, and Recursive Language Models reaching 91.0\% on multi-hop synthesis queries. Finally, a \textbf{\emph{caching layer}} validates query equivalence across multiple dimensions beyond embedding similarity, achieving zero false cache hits and $8.4\times$ latency reduction.
\end{abstract}

\keywords{conversational analytics, NL2SQL, retrieval-augmented generation, semantic caching, query routing, enterprise BI}

\maketitle

\section{Introduction}
\label{sec:intro} 
Enterprise analysts increasingly pose business intelligence questions in natural language, but answers span heterogeneous data: structured warehouses storing transactional metrics and unstructured document repositories containing executive presentations and strategy decks. One analyst asks ``What was the Monthly Active Accounts trend in EMEA last quarter?'' - answerable only from the warehouse. Minutes later, the same analyst asks ``What did the latest strategy deck recommend for EMEA?'' - answerable only from a slide deck. Today, these require different tools; the analyst must decide where the answer lives before they can ask.

Existing AI-enabled BI interfaces address these modalities in isolation. NL2SQL systems~\cite{kumar2024,li2024bird,yu2018spider} translate natural language to SQL but cannot read slide decks. Retrieval-augmented generation pipelines~\cite{gao2024rag_survey,lewis2020rag} excel at document-grounded QA but lack access to live warehouse data. As a result, users must manually determine where an answer resides and invoke the appropriate tool - a cognitive burden that undermines the promise of conversational analytics.
 
We argue that conversational BI should be re-framed as a \emph{heterogeneous query processing problem}. Just as federated query engines perform planning and optimization over distributed relational sources, a conversational BI engine must plan over modality-heterogeneous sources with fundamentally different access methods, cost characteristics, and correctness semantics. This framing introduces challenges absent from any single-modality system: (i)~routing queries to the appropriate modality and retrieval strategy without user intervention, (ii)~adapting retrieval complexity to query difficulty rather than applying a single pipeline uniformly, and (iii)~avoiding redundant computation across semantically equivalent queries while guaranteeing that cached answers are never silently wrong.
 
This paper presents COGNI, a deployed conversational BI system organized as a four-layer pipeline, where each layer addresses a distinct challenge in heterogeneous enterprise analytics:
\begin{enumerate}
\item \textbf{Indexing layer: Slide-adaptive document representation.} Enterprise slide decks encode a three-level hierarchy (deck, slide, typed content blocks) where tables, charts, and narrative carry distinct semantics. We present a slide-adaptive chunking strategy that selects the approach based on content type: recursive chunking for plain-text slides, hierarchical chunking for slides containing structured elements. Combined with dual-index construction over extracted entities and keywords, slide-adaptive chunking lifts Hybrid RAG from 78.3\% to 88.3\% on a 500-question enterprise benchmark (Table~\ref{tab:retrieval-full}); Figure~\ref{fig:cogni-flow} traces the full retrieval ablation that motivated this design.
 
\item \textbf{Routing layer: Dual-output SLM classifier.} The boundary between SQL-answerable and document-answerable queries is subtle, and within the document path, query complexity determines whether standard retrieval or multi-hop reasoning is required. We present a LoRA fine-tuned 1.5B-parameter router trained on a synthetically generated, LLM-validated corpus of 2,812 questions that produces a dual output: a modality decision (SQL vs.\ document) and a complexity assessment selecting the downstream retrieval strategy. The fine-tuned router outperforms frontier-model zero-shot routing at a fraction of the inference cost (Section~\ref{sec:router}, Figure~\ref{fig:rm}).
 
\item \textbf{Retrieval layer: Complexity-adaptive retrieval.} For structured data, a self correcting NL2SQL agent built on LangGraph decomposes SQL generation into discrete stages with two-tier error correction - in-place syntax repair and full query regeneration with accumulated error context - recovering queries that would otherwise propagate as hard failures (Section~\ref{sec:nl2sql}). For unstructured data, the router's complexity signal dispatches queries to one of two strategies: Hybrid RAG with RRF fusion and two-stage LLM reranking for factual queries, or Recursive Language Models for multi-hop synthesis requiring reasoning across entire reports. The complexity-adaptive approach closes the gap on hard-tier questions where standard retrieval is insufficient (Section~\ref{sec:retrieval}).
 
\item \textbf{Caching layer: Verified semantic cache.} Semantic caching over enterprise BI queries is unsafe when gated on embedding similarity alone, because queries differing only in a time period, polarity, or analytical intent produce near-identical embeddings. We present a multi-dimensional equivalence verification model that treats embedding similarity as a necessary but insufficient pre-filter, eliminating the false cache hits that threshold-only caches admit while preserving meaningful recall and latency benefits (Section~\ref{sec:cache}, Table~\ref{tab:cache}).
 
\end{enumerate}
 
We additionally describe deployed system components - vocabulary normalization and multi-turn session management - as system context (Section~\ref{sec:system_context}), not as research contributions, because they are not evaluated independently. Section~\ref{sec:related} discusses related work; Section~\ref{sec:limitations} details limitations; Section~\ref{sec:conclusion} concludes with open problems.

\section{System Architecture}
\label{sec:arch}

Figure~\ref{fig:arch} shows the full architecture. A natural-language query flows through the system top-to-bottom:

\begin{figure*}[!htbp]
    \centering
    \includegraphics[width=0.85\linewidth]{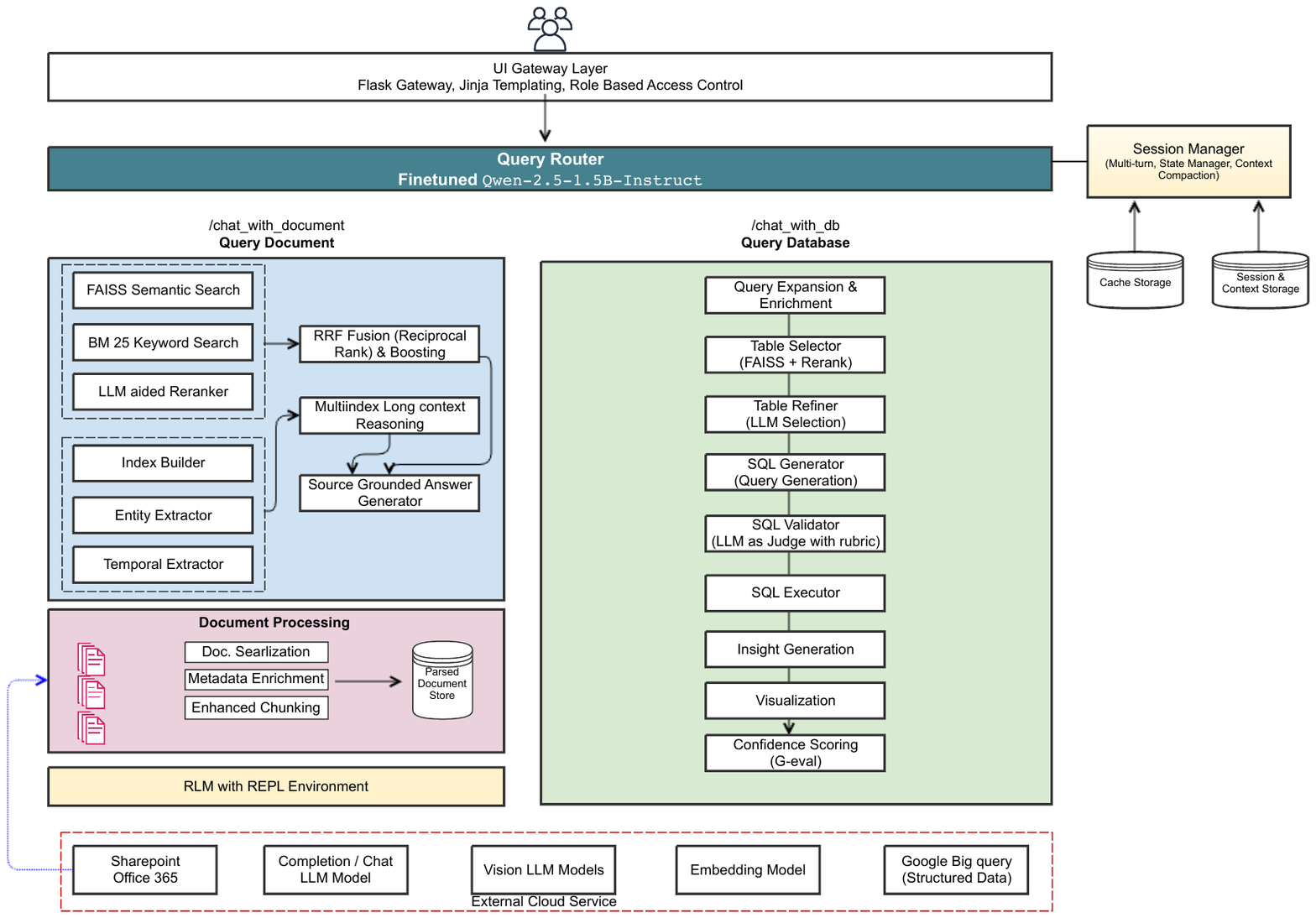}
    \caption{High-level architecture of \textsc{COGNI}.}
    \label{fig:arch}
\end{figure*}

\begin{enumerate}
\item \textbf{Ingress gateway.} Flask endpoint with role-based access control terminates the user session -- isolates auth and tenancy from downstream logic.
\item \textbf{Session manager.} A LangGraph~\cite{langgraph_framework} checkpointer persists conversation history and rehydrates it on each turn -- letting follow-ups like \textit{``what about Germany?''} inherit the prior metric, region, and timeframe.
\item \textbf{Vocabulary normalizer.} A synonym registry (Section~\ref{sec:vocab_brief}) rewrites the query to canonical surface forms -- guarantees the cache, retrieval, and NL2SQL layers all see the same entity strings.
\item \textbf{Verified semantic cache (Section~\ref{sec:cache}).} Conjunctive checks on embedding similarity, entity overlap, intent, polarity, and freshness gate a cached response -- short-circuits redundant pipeline work while preventing the false hits that embedding-only caches admit.
\item \textbf{SLM router (Section~\ref{sec:router}).} A LoRA-fine-tuned 1.5B classifier emits a modality decision (SQL vs.\ document) and a complexity signal -- routes the query to one pipeline rather than running both, avoiding redundant-latency fallback.
\item \textbf{NL2SQL pipeline (Section~\ref{sec:nl2sql}).} For SQL-bound queries, a graph-structured agent runs table selection $\rightarrow$ SQL generation $\rightarrow$ validation $\rightarrow$ execution over BigQuery~\cite{google_bigquery}, with two-tier self-correction on failure -- recovers queries that would otherwise propagate column or type errors as hard failures.
\item \textbf{Document pipeline (Section~\ref{sec:retrieval}).} For document-bound queries, the router's complexity signal selects either Hybrid RAG with RRF~\cite{cormack2009reciprocal} fusion (factual queries) or a Recursive Language Model~\cite{rlm} path (multi-hop synthesis) -- trades latency for reasoning depth only when the query needs it.
\item \textbf{Indexing layer (offline, feeds the document pipeline).} SharePoint~\cite{microsoft_sharepoint_collaboration} sync $\rightarrow$ vision-LLM extraction $\rightarrow$ slide-adaptive chunking $\rightarrow$ dual FAISS~\cite{faiss_library} + BM25~\cite{robertson2009bm25} indexes -- exposes a single retrieval interface regardless of chunking strategy, keeping the retrieval layer agnostic to upstream choices.
\item \textbf{Answer generation.} The retrieved evidence is passed to an answer LLM with citation requirements, and the response is written back to the cache before returning to the user -- closes the loop so the next equivalent query is served from cache.
\end{enumerate}

Steps 1--3 prepare the query, step 4 short-circuits redundant work, steps 5--7 perform the heterogeneous retrieval planning that motivates the system, step 8 supplies the corpus that step 7 reads from, and step 9 closes the loop. Sections~\ref{sec:retrieval}--\ref{sec:cache} describe the four contribution layers in detail.

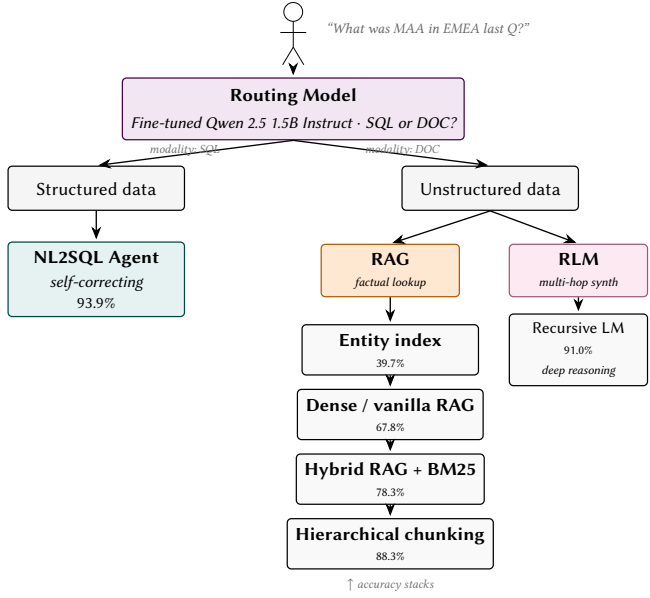
\begin{figure}[t]
\centering
\resizebox{\linewidth}{!}{%
\begin{tikzpicture}[
  >={Stealth[length=2.2mm, width=1.7mm]},
  font=\sffamily,
  box/.style={
    draw,
    line width=0.6pt,
    rounded corners=2pt,
    align=center,
    inner sep=4pt
  },
  router/.style={
    box,
    fill=violet!10,
    draw=violet!55!black,
    minimum width=4.3cm,
    minimum height=1.05cm
  },
  data/.style={
    box,
    fill=gray!8,
    minimum width=3.0cm,
    minimum height=0.75cm
  },
  nl2sql/.style={
    box,
    fill=teal!10,
    draw=teal!55!black,
    minimum width=3.0cm,
    minimum height=1.15cm
  },
  rag/.style={
    box,
    fill=orange!18,
    draw=orange!70!black,
    minimum width=2.35cm,
    minimum height=0.85cm
  },
  rlm/.style={
    box,
    fill=magenta!10,
    draw=magenta!60!black,
    minimum width=2.35cm,
    minimum height=0.85cm
  },
  stage/.style={
    box,
    fill=gray!5,
    minimum width=3.0cm,
    minimum height=0.75cm
  },
  rlmd/.style={
    box,
    fill=gray!5,
    minimum width=2.35cm,
    minimum height=1.05cm
  },
  arr/.style={
    ->,
    line width=0.65pt,
    shorten >=1pt
  },
]

\begin{scope}[shift={(0,8.8)}, line width=0.45pt]
  \draw (0,0.3) circle (0.16);
  \draw (0,0.14) -- (0,-0.30);
  \draw (-0.22,-0.05) -- (0.22,-0.05);
  \draw (0,-0.30) -- (-0.18,-0.60);
  \draw (0,-0.30) -- (0.18,-0.60);
\end{scope}

\node[font=\itshape\footnotesize, color=black!60, anchor=west]
  at (0.45, 8.8) {``What was MAA in EMEA last Q?''};

\node[router] (router) at (0, 7.45)
  {{\bfseries Routing Model}\\[2pt]
   {\small\itshape Fine-tuned Qwen 2.5 1.5B Instruct $\cdot$ SQL or DOC?}};

\draw[arr] (0, 8.15) -- (router.north);

\node[data] (struct)   at (-3.35, 6.1) {Structured data};
\node[data] (unstruct) at ( 3.35, 6.1) {Unstructured data};

\draw[arr] (router.south) -- (struct.north);
\draw[arr] (router.south) -- (unstruct.north);

\node[font=\scriptsize\itshape, color=black!55]
  at (-1.85, 6.75) {modality: SQL};

\node[font=\scriptsize\itshape, color=black!55]
  at (1.85, 6.75) {modality: DOC};

\node[nl2sql] (nl2sql) at (-3.35, 4.55)
  {{\bfseries NL2SQL Agent}\\[1pt]
   {\small\itshape self-correcting}\\[-1pt]
   {\small 93.9\% }};

\node[rag] (rag) at (1.65, 4.7)
  {{\bfseries RAG}\\[-1pt]
   {\scriptsize\itshape factual lookup}};

\node[rlm] (rlm) at (4.85, 4.7)
  {{\bfseries RLM}\\[-1pt]
   {\scriptsize\itshape multi-hop synth}};

\draw[arr] (struct.south) -- (nl2sql.north);
\draw[arr] (unstruct.south) -- (rag.north);
\draw[arr] (unstruct.south) -- (rlm.north);

\node[stage] (v1) at (1.65, 3.35)
  {{\bfseries Entity index}\\[-1pt]
   {\scriptsize 39.7\%}};

\node[stage] (v2) at (1.65, 2.25)
  {{\bfseries Dense / vanilla RAG}\\[-1pt]
   {\scriptsize 67.8\%}};

\node[stage] (v4) at (1.65, 1.15)
  {{\bfseries Hybrid RAG + BM25}\\[-1pt]
   {\scriptsize 78.3\%}};

\node[stage] (hi) at (1.65, 0.05)
  {{\bfseries Hierarchical chunking}\\[-1pt]
   {\scriptsize 88.3\%}};

\draw[arr] (rag.south) -- (v1.north);
\draw[arr] (v1.south) -- (v2.north);
\draw[arr] (v2.south) -- (v4.north);
\draw[arr] (v4.south) -- (hi.north);

\node[font=\scriptsize\itshape, color=black!55]
  at (1.65, -0.65) {$\uparrow$ accuracy stacks};

\node[rlmd] (rlmd) at (4.85, 3.35)
  {{\small Recursive LM}\\[-1pt]
   {\scriptsize 91.0\%  }\\[-1pt]
   {\scriptsize\itshape deep reasoning}};

\draw[arr] (rlm.south) -- (rlmd.north);

\end{tikzpicture}%
}

\caption{COGNI query flow. A query enters the routing model, which emits a modality decision (SQL vs.\ document) and, for the document path, a complexity signal selecting between Hybrid RAG and Recursive LM.}
\label{fig:cogni-flow}
\end{figure}

The routing model (step 5) is the architectural hinge: everything upstream of it -- gateway, session, vocabulary, cache -- prepares a normalized query without committing to a pipeline, and everything downstream is selected by what the router emits. Figure~\ref{fig:cogni-flow} traces this decision end-to-end. A LoRA-fine-tuned Qwen 2.5 1.5B Instruct classifier (Section~\ref{sec:router}) produces a dual output on each query: a \emph{modality decision} that picks between the NL2SQL agent (Section~\ref{sec:nl2sql}, 93.9\% G-Eval) and the document path, and a \emph{complexity signal} that, within the document path, picks between Hybrid RAG for factual lookups and Recursive Language Models for multi-hop synthesis (Section~\ref{sec:retrieval}). Routing replaces the document-first-then-SQL fallback that an undirected system would default to SQL-answerable queries no longer pay document-retrieval latency, and the expensive RLM path is invoked only when the complexity signal warrants it.


\section{Document Retrieval Pipeline}
\label{sec:retrieval}

The document journey has two stages: an \emph{indexing} stage that converts enterprise slide decks into a queryable knowledge base, and a \emph{retrieval} stage that, for each query, selects from two strategies based on the router's complexity signal. Both stages operate over the same corpus and feed the same answer generator. The design choices below were arrived at through an internal ablation on a 500-question enterprise benchmark spanning easy (single-slide), medium (multi-page), and hard (multi-deck) complexity tiers, evaluated with a weighted G-Eval~\cite{geval} rubric (factual correctness 40\%, completeness 35\%, relevance 25\%, 0--10 scale). Table~\ref{tab:retrieval-full} reports the full ablation; the indexing and retrieval subsections that follow describe each design choice in the order it was added.

\begin{table}[!htbp]
\small
\centering
\begin{tabular}{lc}
\toprule
\textbf{Configuration} & \textbf{G-Eval (\%)} \\
\midrule
Entity-overlap indexing only            & 39.7 \\
+ Dense semantic RAG                    & 67.8 \\
+ BM25 sparse index with RRF fusion     & 78.3 \\
+ Slide-adaptive hierarchical chunking  & \textbf{88.3} \\
\midrule
Recursive Language Models (alt.\ retrieval) & \textbf{91.0} \\
\bottomrule
\end{tabular}
\caption{Document retrieval on the 500-question enterprise benchmark. Rows 1--4 are cumulative indexing/retrieval design choices; the deployed default is row 4 (88.3\%). Recursive Language Models (row 5) is an alternative retrieval strategy selected by the router for multi-hop synthesis queries, not a further stage on top of row 4.}
\label{tab:retrieval-full}
\end{table}

\subsection{Indexing}

Indexing converts a deck into three parallel representations that downstream retrieval can combine: a structural hierarchy, a dense embedding index, and a sparse keyword index. We describe each in the order it was added to the production system; cumulative impact is reported in Table~\ref{tab:retrieval-full}.

\paragraph{Entity-overlap baseline (39.7\%).}
The earliest production indexing was a structured-entity baseline: a vision-LLM extracts KPIs, regions, products, merchants, and temporal references at ingestion time, normalizes to canonical forms via the synonym registry (Section~\ref{sec:vocab_brief}), and inserts each entity type into a dedicated inverted index. Query-time retrieval scored slides via weighted entity overlap (KPI=3.0, region/product=2.0, merchant=1.5, with temporal boost). At 39.7\% on the 500-question benchmark, this confirmed that structured lookup alone cannot handle the breadth of free-form or multi-hop questions; subsequent design choices stacked semantic retrieval on top.

\paragraph{Dense embedding index (+28.1 pp $\rightarrow$ 67.8\%).}
Slides are chunked at the 2{,}500-token boundary with 200-token overlap, each block embedded with \texttt{text-embedding-3-large} and indexed in FAISS~\cite{faiss_library} for cosine-similarity retrieval. Dense retrieval contributed the largest single gain in the ablation. Lexical overlap with free-form enterprise queries is weak (``what drove the decline'' rarely contains the exact words used in the source slide), making semantic retrieval the dominant signal.

\paragraph{Sparse keyword index with RRF fusion (+10.5 pp $\rightarrow$ 78.3\%).}
A BM25~\cite{robertson2009bm25} index is built in parallel over each chunk's \emph{extracted keywords and entities} (not raw text). Dense and sparse signals are fused via Reciprocal Rank Fusion (Section~\ref{sec:retrieval-strategies}), recovering precision on KPI lookups where exact acronym match matters (GNA, TPV, NPS) -- a regime where dense retrieval underperforms.

\paragraph{Slide-adaptive hierarchical chunking (+10.0 pp $\rightarrow$ 88.3\%).}
Flat token-boundary chunking destroys structural narrative: a chunk from slide 12 has no awareness of what deck it belongs to or what section the slide is in. We index documents along the natural hierarchy following HiChunk-style~\cite{hichunk} segmentation -- L1~deck, L2~slide, L3~typed leaf blocks (text, table, chart, key-value). Tables and charts carry both a natural-language summary (for retrieval embedding) and full structured content (for the answer generator). When multiple L3 blocks from the same slide are retrieved at query time, the full L2 slide is auto-merged into context, preserving structural narrative that flat chunking strips out. This is the deployed default.

\subsection{Retrieval}
\label{sec:retrieval-strategies}

Within the document path, the router (Section~\ref{sec:router}) selects between two retrieval strategies based on its complexity signal. Both write into the same answer-generation interface, so the downstream LLM is agnostic to which strategy ran.

\paragraph{Hybrid RAG with RRF fusion.}
The default strategy for factual queries, operating over the indexes described above. Dense and sparse retrieval execute in parallel: FAISS retrieves $2 K_{\text{init}}$ candidates by cosine similarity, BM25 retrieves $K_{\text{init}}$ candidates by lexical overlap on extracted entities ($K_{\text{init}} = 40$). Scores fuse via Reciprocal Rank Fusion~\cite{cormack2009reciprocal}:
\[
\text{RRF}(i) = \frac{w_s}{k + r_s(i)} + \frac{w_k}{k + r_k(i)},
\]
with $k = 60$, $w_s = 0.70$, $w_k = 0.30$. RRF is score-agnostic, sidestepping the incompatible scales of cosine similarity and BM25. When dense retrieval returns low-confidence results, a fallback policy promotes BM25 candidates, recovering KPI-acronym queries that the dense index handles poorly. Top fused candidates feed two-stage LLM reranking: relevance scoring (0--10) followed by factual alignment scoring (0--10) on the top-15, with final score $0.6 \cdot s_{\text{rel}} + 0.4 \cdot s_{\text{fact}}$.

\paragraph{Recursive Language Models (91.0\%).}
The alternative strategy for multi-hop synthesis queries -- questions that require reasoning across an entire report rather than reading a fixed retrieval window. Recursive Language Models~\cite{rlm} reframe the document as an external environment, not as prompt content: a root LM emits code that decomposes the document programmatically, invokes a sub-LM over slices, and accumulates results in a REPL until it can answer. With \texttt{max\_depth=30} and \texttt{max\_calls\_per\_subagent=20}, RLM trades latency for reasoning depth -- a complex query may trigger 15--25 sub-LM calls. Routing to RLM happens only when the complexity signal warrants it; the two strategies are complementary alternatives, not stacked, with Hybrid RAG recovering structural narrative on easy and medium tiers and RLM addressing the hard tier where retrieval-then-read is fundamentally insufficient.

\section{Self-Correcting NL2SQL Agent}
\label{sec:nl2sql}

Enterprise warehouses contain hundreds of tables with column identifiers that diverge systematically from natural-language references. A query asking for ``average order value in EMEA last quarter'' must resolve to
\begin{center}
\small\texttt{SUM(total\_payment\_volume)\,/\,SUM(transaction\_count)}
\end{center}
filtered to a region encoded as one of \{EU, EMEA, MEA, EUROPE\} depending on the table. Frontier LLMs generate SQL that is often syntactically correct but semantically wrong~\cite{pourreza2024dinsql, gao2024dailsql}. We address this with a graph-structured agent built on LangGraph that decomposes NL2SQL into discrete stages with self-correcting retry loops at two granularities.

\paragraph{Agent architecture.}
The pipeline is modeled as a directed state graph $G = (V, E)$ over six nodes (\texttt{table\_sel}, \texttt{sql\_gen}, \texttt{sql\_val}, \texttt{query\_exec}, \texttt{insights\_gen}, \texttt{error\_handler}), with shared state $S = \langle q, T, \sigma, \mathcal{E}, r, R \rangle$ (query, selected tables, generated SQL, error history, retry counter, results). The error\_handler is reachable from conditional edges when retries are exhausted ($r \geq 3$).

\paragraph{Schema linking.}
A curated YAML metadata layer captures, per table: measures (NL synonyms, column mappings via \texttt{db\_mapping}, aggregation formulas) and dimensions (type, enumerated allowed values). The table selector prompts an LLM with a compact schema summary; selected tables are expanded into full metadata for SQL generation. The validator applies rule-based checks without LLM invocation: the query must begin with \texttt{SELECT} or \texttt{WITH}; no destructive DDL keywords; referenced columns must exist in selected table metadata.

\paragraph{Two-tier error correction.}
On execution failure, the agent distinguishes between two error classes. \emph{Surface-level syntax errors} (missing commas, unclosed parentheses, invalid function names, column typos) are handled by a query-fixer prompt that attempts up to 2 in-place fixes \emph{without consuming a graph-level retry}. \emph{Semantic errors} (column not found, type mismatch, missing filter) append the error to $\mathcal{E}$, increment $r$, and return control to \texttt{sql\_gen} in error-correction mode where the full error history informs the next attempt. Before execution, a smart \texttt{LIMIT} policy removes the \texttt{LIMIT} clause for aggregated queries (detected via \texttt{GROUP BY} or aggregate functions) and preserves \texttt{LIMIT 1000} otherwise.

\paragraph{Evaluation.}
On 100 hand-curated golden (NL query, SQL) pairs covering trends, segmentation, year-over-year growth, seasonal patterns, root-cause analysis, and country-level ranking, the agent achieves 93.9\% G-Eval~\cite{geval} score on three weighted dimensions: SQL relevance (40\%), table selection (35\%), query understanding (25\%). Per-dimension: table selection 97.8\%, query understanding 99.2\%, SQL relevance 84.6\%. Hard-fail rules cap correctness scores for structural errors (wrong table $\leq 4$, missing mandatory filters $\leq 5$, non-existent columns $\leq 6$). SQL relevance is the hardest dimension, driven by complex filter clauses and multi-column aggregation formulas requiring exact formula application from metadata. The two-tier correction strategy is responsible for recovering queries that fail on the first generation; without it, execution failures due to column mismatches and type errors would propagate as hard failures.

\section{Fine-Tuned SLM Router}
\label{sec:router}

\textsc{COGNI}'s two retrieval paths answer different question shapes. Quantitative queries with aggregations or temporal scope (\textit{``What is the total TPV for US Casual Buyers in December 2025?''}) belong on the SQL path; qualitative queries seeking strategic context or definitions (\textit{``According to the April 9th CFS roadmap, what are the new launch dates?''}) belong on the document path. A naive document-first-then-SQL fallback adds redundant latency on every SQL-answerable query. The boundary is often subtle: \textit{``What is AOV?''} reads like SQL because AOV is a metric, but without temporal or dimensional context it is a definitional query best answered from a deck. Rule-based heuristics struggle with this distinction; we present a learned classifier.

\paragraph{Synthetic training data.}
Manual annotation at scale was impractical, so we built a four-stage synthetic data pipeline grounded in 12 schema YAMLs, 35 weekly business review decks, and 150 hand-labeled seed question-route pairs. \emph{Pattern extraction} derived 8 indicators per route from the seeds (SQL: metrics, calculations, temporal scope, ranking; DECK: explicit slide references, qualitative insights, ownership, context-free definitional queries). \emph{Question generation} used gemini-2.5-flash and claude-sonnet-4-6 to independently produce $\sim$2{,}000 questions each across three complexity tiers; each batch was grounded in one YAML and one DECK file, cycled through all 47 source files, with 15 SQL and 15 DECK style seeds rotated per batch. Inline deduplication via fuzzy string matching (sequence-matcher threshold 0.85) caught near-duplicates. \emph{Classification} labeled each question conditioned on the indicators at temperature 0.1. \emph{Multi-stage validation} scored every candidate 1--5 with claude-opus-4-6 (with a 97\% token reduction by loading only matched YAML+DECK files); only candidates scoring $\geq 4$ entered the final dataset. A 100-question manual audit yielded 91\% labelling accuracy. Final dataset: 2{,}812 questions (1{,}169 SQL, 1{,}643 DECK), split 80/20 with stratification on route and complexity, yielding 2{,}249 training and 563 test rows.

\paragraph{Model and training.}
We fine-tuned Qwen 2.5 1.5B Instruct~\cite{qwen25} using LoRA~\cite{lora} via the Unsloth framework~\cite{unsloth}, selected over Gemma 3 1B Instruct~\cite{gemma3} after a sample-testing phase. LoRA adapters are injected into all 7 target modules (4 attention projections plus 3 MLP projections) with rank $r=32$, $\alpha=64$. The choice of rank $(r)$ and $\alpha$ is recommended by \cite{rathore-etal-2025-much} and \cite{biderman2024lora}. Only 36.9M parameters are trainable (2.34\% of 1.58B total); base weights remain frozen. Training: learning rate $5 \times 10^{-5}$ with linear scheduling, 3 epochs, effective batch size 64 (per-device 4 with gradient accumulation 16). The model receives each question wrapped in Qwen's chat template; expected output is tuple of (\texttt{SQL} or \texttt{DECK}) and complexity, avoiding parsing overhead. Loss converged steadily (train 1.5451 $\rightarrow$ 1.0838; eval 1.4961 $\rightarrow$ 1.0971 over 3 epochs).

\begin{figure}[h!]
    \centering
    \includegraphics[width=\linewidth]{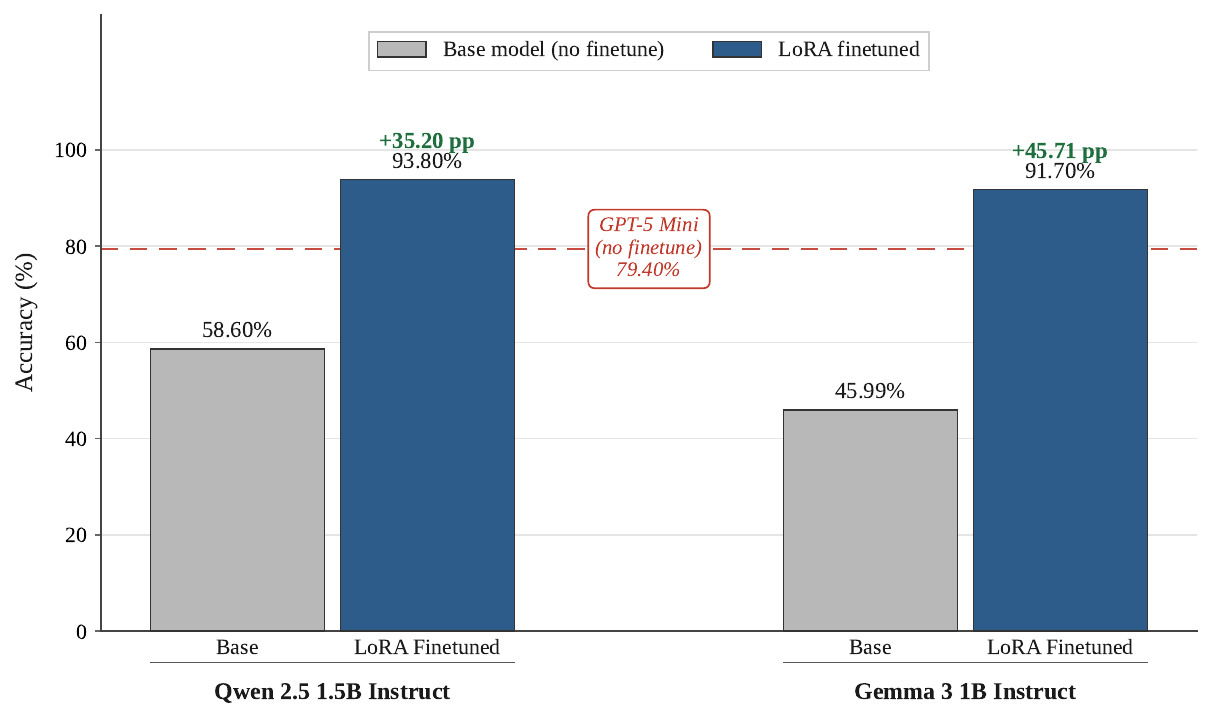}
    \caption{Routing accuracy  on Finetuned Model}
    \label{fig:rm}
\end{figure}

\paragraph{Evaluation.}
On 563 held-out test questions, the fine-tuned router achieves 93.8\% accuracy (528 / 563). Figure~\ref{fig:rm} compares against three baselines under identical prompts and generation parameters.


\paragraph{Cost profile.}
The fine-tuned model runs on a single T4 GPU at a fixed monthly cost of \$360 (\$0.50/hour $\times$ 720 hours) independent of request volume. GPT-5-Mini's per-token pricing scales linearly: $\sim$\$2.50/month at 10K requests, $\sim$\$2{,}500/month at 10M requests. At enterprise scale, the fine-tuned model is approximately $7\times$ cheaper at higher accuracy. Single-pass inference produces a routing decision in under 100\,ms on the T4. The 35.2 pp gap between fine-tuned and base Qwen demonstrates that the routing decision is not learnable in-context from a system prompt alone -- domain-specific fine-tuning on synthetically generated, validated data is what closes it.

\section{Verified Semantic Cache}
\label{sec:cache}

Semantic caching for LLM-backed systems typically gates cache hits on embedding similarity~\cite{bang2023gptcache}. In enterprise BI this is unsafe: queries that differ only in a time period (\emph{``MAU in Q1 2024''} vs.\ \emph{``MAU in Q1 2025''}), polarity (\emph{``revenue growth''} vs.\ \emph{``revenue decline''}), or analytical intent (lookup vs.\ comparison) often exceed 0.95 cosine similarity under \texttt{text-embedding-3-large}, producing silent factual errors. A cache that returns false hits is strictly worse than no cache, since users receive confidently-presented wrong data. Our design principle is therefore asymmetric: cache misses are tolerable; false hits are not.

\paragraph{Multi-dimensional equivalence verification.}
Rather than relying on a single similarity score, we define a cache hit as the conjunction of multiple independent equivalence checks, each enforcing a distinct dimension of query semantics. Embedding similarity serves as a necessary but insufficient pre-filter; subsequent checks apply deterministic, domain-grounded verification before a cached response is served. The pipeline follows a \emph{fail closed} design: failure at any check produces an immediate cache miss, and a hit is returned only when all checks pass. Checks are ordered by increasing computational cost to minimize average verification latency on rejected queries.

The verification dimensions are informed by an empirical analysis of failure modes in enterprise BI queries. We identify four classes of semantically distinct queries that embedding models conflate: \emph{entity substitution} (different KPI, region, or time period), \emph{intent variation} (e.g., lookup vs.\ comparison vs.\ trend analysis), \emph{polarity reversal} (e.g., growth vs.\ decline), and \emph{temporal staleness} (cached answers to queries about ``current'' or ``latest'' data that have expired). Each class requires a dedicated verification mechanism; no single check subsumes the others.

\paragraph{Evaluation.}
We evaluate on 200 queries expanded from 50 hand-curated seeds across five categories: paraphrase hits (54 from 12 seeds), entity mismatches (62 from 15), intent mismatches (36 from 10), negation reversals (30 from 8), and edge cases (18 from 5). Results are summarized in Table~\ref{tab:cache}.

\begin{table}[h]
\centering
\caption{Cache effectiveness on 200 evaluation queries.}
\label{tab:cache}
\begin{tabular}{lr}
\toprule
\textbf{Metric} & \textbf{Value} \\
\midrule
Cache hits (true positives) & 47\,/\,54 expected (87.0\% recall) \\
Cache misses (true negatives) & 146\,/\,146 expected (100\%) \\
False cache hits & 0 \\
False cache misses & 7 \\
Precision & 1.000 \\
Avg.\ latency -- cache hit path & 408\,ms \\
Avg.\ latency -- cache miss path & 3{,}420\,ms \\
Cache hit speedup & 8.4$\times$ \\
\bottomrule
\end{tabular}
\end{table}

The cache produces \textbf{zero false hits} across all 200 queries. The 7 false misses are paraphrase queries whose embedding similarity fell below the pre-filter threshold despite semantic equivalence; these proceed harmlessly through the full pipeline. The 87.0\% recall alongside 100\% precision reflects the asymmetric design principle stated above. Critically, every verification dimension contributes non-zero rejections that no other dimension catches: an embedding-only cache at the same similarity threshold would admit all 146 negative queries as false hits, since every rejected query passes the embedding pre-filter. The multi-dimensional verification overhead ($\sim$20\,ms beyond embedding computation) is justified by the correctness guarantee it provides. On the cache-hit path, average latency is 408\,ms (dominated by the embedding API call), an 8.4$\times$ speedup over the full pipeline (3{,}420\,ms). In production workloads with observed hit rates of 35 -40\%, the projected blended latency reduction is 30--35\%.

\section{System Context}
\label{sec:system_context}

\paragraph{Vocabulary normalization.}
\label{sec:vocab_brief}
A synonym registry maps domain concepts to canonical surface forms across six entity types (metrics, temporal references, regions, product lines, segments, organizational units), with over 150 entries. At query time, the normalizer performs a single left-to-right pass with greedy longest-match using $n$-gram probes for $n \in \{4, 3, 2\}$, plus single-token lookups. Two-character tokens are expanded only when fully uppercase, avoiding collisions with common English words. The pass is $O(n)$ in query tokens with sub-millisecond overhead. Cache correctness (Section~\ref{sec:cache}) and retrieval recall depend on uniform normalization across modalities.

\paragraph{Multi-turn session management.}
Sessions are identified by UUID with state persisted via a LangGraph checkpointer over SQLite, replicated asynchronously to BigQuery. A context-window hook trims history to the most recent ten turns. A finite-state machine governs flow across discrete nodes (\texttt{START}, \texttt{DECK}, \texttt{SQL}, \texttt{CONSTRAINED}, \texttt{KNOW\_MORE}, \texttt{END}), enabling follow-up resolution (\textit{``What about Germany?''} inheriting prior metric and timeframe).

\section{Related Work}
\label{sec:related}

\paragraph{Retrieval-augmented generation.}
Hybrid retrieval combining dense vectors with BM25~\cite{robertson2009bm25} via Reciprocal Rank Fusion~\cite{cormack2009reciprocal} is the standard recipe. Li et al.~\cite{li2025ragbestpractices} provide a systematic study of RAG best practices. HiChunk~\cite{hichunk} argues for hierarchical chunking; Recursive Language Models~\cite{rlm} propose iterative evidence-driven retrieval for multi-hop queries; PageIndex~\cite{pageindex} introduces vectorless reasoning-based RAG. Our work evaluates these approaches under identical conditions on a single enterprise benchmark.

\paragraph{NL2SQL}
Spider~\cite{yu2018spider} and BIRD~\cite{li2024bird} established cross-domain NL2SQL benchmarks. CHESS~\cite{talaei2024chess} and CHASE-SQL~\cite{pourreza2025chasesql} advanced accuracy through multi-agent pipelines and preference-optimized candidate selection. DAIL-SQL~\cite{gao2024dailsql} focuses on few-shot retrieval-augmented prompting. These systems operate over structured tables only. Our agent is simpler than CHASE-SQL -- single-path generation with two-tier error correction rather than multi-path with selection -- targeting a different regime: enterprise schemas where the bottleneck is vocabulary mismatch rather than reasoning depth.

\paragraph{Query routing and SLM fine-tuning.}
ReAct-style agents~\cite{yao2023react} pick tools via in-context reasoning, and recent work~\cite{seabra2024multiagent} explores multi-agent orchestration over heterogeneous sources. We treat routing as a discriminative classification task amenable to fine-tuning a small model on a synthetically generated, validated corpus, following LoRA~\cite{lora} on sub-2B base models (Qwen 2.5 1.5B~\cite{qwen25}, Gemma 3 1B~\cite{gemma3}). The result echoes findings in domain-specific text classification: small specialist models outperform frontier generalists on focused tasks.

\paragraph{Semantic caching.}
GPTCache~\cite{bang2023gptcache} is the canonical reference for LLM-backed semantic caching. Its core design - embed, retrieve nearest neighbors, return response if similarity exceeds threshold - is shared by subsequent systems. None verifies equivalence beyond similarity. COGNI's verified cache adds deterministic, domain-grounded equivalence checks beyond embedding similarity, addressing failure modes - entity substitution, intent
variation, polarity reversal, temporal staleness - that threshold-only caches cannot detect.

\paragraph{LLM-as-judge.}
G-Eval~\cite{geval} and subsequent surveys~\cite{gu2024llmasjudge} establish LLM-as-judge as a scalable evaluation paradigm. We use it for reranking within the retrieval pipeline and for end-to-end scoring.

\section{Limitations}
\label{sec:limitations}

\paragraph{Internal benchmarks.}
Our experimental results are evaluated and reported on internal enterprise benchmarks.  We should evaluate our system on public benchmarks such as Spider \cite{yu2018spider}, BIRD\cite{li2024bird}, MultiModalQA or MultiHop-RAG.

\paragraph{Cache evaluation construction.}
The 200-query cache set is constructed adversarially: it deliberately oversamples queries that an embedding-only cache would mis-accept (entity substitutions, polarity reversals, intent shifts). Our zero-false-hit result therefore demonstrates correctness on the population our cache was designed to defend against, not on arbitrary production traffic. We observe that an embedding-only cache at a comparable similarity threshold would admit all 146 negatives as false hits, since every rejected query passes the embedding pre-filter; we have not run GPTCache as a direct baseline.

\paragraph{NL2SQL evaluation.}
The 93.9\% G-Eval score on 100 enterprise queries is not directly comparable to execution accuracy on Spider-dev or BIRD. Our benchmark uses a rubric-grounded scoring approach because correct answers are non-deterministic under live warehouse state, but this means we cannot claim our agent would beat CHASE-SQL~\cite{pourreza2025chasesql} or DAIL-SQL~\cite{gao2024dailsql} on their reported metrics. The agent targets a different regime -- enterprise schemas with vocabulary mismatch rather than reasoning depth -- and the high score should be read as evidence that two-tier self-correction handles this regime, not as a SOTA claim.

\paragraph{Router generalization.}
The router is trained on 2{,}814 synthetic question-route pairs generated by gemini-2.5-flash and claude-sonnet-4-6, validated by claude-opus-4-6. The 91\% labelling accuracy on a 100-question human audit gives confidence in the training signal but does not bound the model's accuracy on real production traffic over time. Class drift -- a gradual shift in the SQL-to-DECK ratio or in question phrasing -- is mitigated by periodic re-validation against the live endpoints, but a formal monitoring framework is future work.

\paragraph{Model-as-judge bias.}
G-Eval~\cite{geval} uses GPT-5 as judge; retrieval and SQL stages use claude-sonnet-4-6 (Appendix~\ref{sec:appendix-models}). We have not investigated whether the judge systematically favors outputs from one model family over another. Validating G-Eval scores against human ratings on a held-out sample, which we plan to do in future work, would partially address this for retrieval but not for SQL.

\paragraph{Latency at the RLM extension.}
The 91.0\% retrieval score from Recursive Language Models comes at a per-query cost of 15--25 sub-LM calls. We treat RLM as a high-value path for complex synthesis queries, gated by the router, rather than as the default. The cost-quality trade-off across recursion depth and sub-agent model choice is an open optimization problem.

\section{Conclusion}
\label{sec:conclusion}
We presented \textsc{COGNI}, a deployed conversational BI system with four contributions: a document pipeline whose slide-adaptive chunking lifts Hybrid RAG to 88.3\% on a 500-question enterprise benchmark, with Recursive Language Models extending coverage to 91.0\% on the multi-hop synthesis subset; a self-correcting NL2SQL agent reaching 93.9\% G-Eval on enterprise queries; a LoRA-fine-tuned 1.5B-parameter router classifying queries between SQL and document paths at 93.8\% accuracy ($\sim$7$\times$ cheaper than GPT-5-Mini zero-shot at production scale); and a verified semantic cache with zero false hits on 200 evaluation queries and $8.4\times$ latency reduction on hits. Open problems include cross-modal answer fusion when a question requires both SQL and document evidence, agentic exploration for retrieval failures on hard-tier questions, drift monitoring for the router as production traffic distributions shift, and adapting the verified cache design to domains with different correctness/latency trade-offs.

\bibliographystyle{ACM-Reference-Format}
\bibliography{references}
\balance
\appendix
\section*{Appendix}

\section{Production Model Configuration}
\label{sec:appendix-models}

\noindent
\begin{center}
\small
\resizebox{0.475\textwidth}{!}{%
\begin{tabular}{@{}ll@{}}
\toprule
\textbf{Component} & \textbf{Model} \\
\midrule
\multicolumn{2}{@{}l}{\textit{Retrieval}} \\
\quad Embedding & text-embedding-3-large (3072d) \\
\quad LLM reranking / answer synthesis & claude-sonnet-4-6 \\
\quad Hierarchical extraction (vision) & gemini-2.5-flash \\
\quad RLM root / sub-agent & GPT-5 / GPT-4o \\
\midrule
\multicolumn{2}{@{}l}{\textit{NL2SQL}} \\
\quad Table selection / SQL gen / correction & claude-sonnet-4-6 \\
\midrule
\multicolumn{2}{@{}l}{\textit{SLM Router}} \\
\quad Base model & Qwen 2.5 1.5B Instruct \\
\quad Fine-tuning & LoRA, $r{=}32$, $\alpha{=}64$ \\
\quad Synthetic data generation & gemini-2.5-flash + claude-sonnet-4-6 \\
\quad LLM-as-judge validator & claude-opus-4-6 \\
\midrule
\multicolumn{2}{@{}l}{\textit{Semantic Cache}} \\
\quad Embedding & text-embedding-3-large (3072d) \\
\midrule
\multicolumn{2}{@{}l}{\textit{Evaluation}} \\
\quad G-Eval judge & GPT-5 \\
\bottomrule
\end{tabular}}
\captionof{table}{Production model configuration by component.}
\label{tab:model-config}
    
\end{center}

\section{Key Prompts}
\label{sec:appendix-prompts}
\noindent This appendix contains the prompts supporting the following contributions: NL2SQL generation and error correction (Section~\ref{sec:nl2sql}), the cache entity extractor (Section~\ref{sec:cache}), and the G-Eval rubric used for SQL evaluation. Template variables are shown in \{curly braces\}. Additional prompts (retrieval reranking, response comparison, and synthetic data generation) are available in the supplementary material archive.








\subsection{NL2SQL Generator}

\begin{lstlisting}[style=prompt]
You are given:
  - User Query: "{user_query}"
  - Table Metadata: {table_metadata}

Your task is to generate a SQL query that accurately
answers the user query using only relevant fields and
appropriate tables based on the structure provided.

***CRITICAL COLUMN USAGE RULES***
1. ONLY use column names from <db_mapping> tags in metadata
2. DO NOT use measure names directly (e.g., "AOV", "ARPA")
   - these are labels, not columns
3. For calculated metrics:
   - Check if there's a <formula> tag - use that exact formula
   - Otherwise, derive from basic columns (e.g., TPV/Txns
     for AOV)
4. For column names with spaces or special characters:
   - Use backticks: `column name with spaces`

***CRITICAL REQUIREMENTS***
- Check table descriptions for mandatory filter requirements
- Pay attention to dimension definitions that mention
  "Always" or "must" requirements
- Do not assume or hardcode values for any columns
- Prefer subqueries or filtering from actual table data

***Column Selection Strategy***
1. First, identify what columns exist in <db_mapping> tags
2. For aggregated metrics mentioned in user query:
   - Check if column exists in db_mapping
   - If not, check for <formula> in measures section
   - If no formula, calculate from base columns
3. Always validate column names match exactly

Give only the SQL as output, no explanations.
\end{lstlisting}

\subsection{NL2SQL Error Correction}

Invoked when the previous SQL execution fails. Receives the failed SQL, the error message, and the schema. Up to two in-place attempts are made before the error escalates to a full regeneration with accumulated error context.

\begin{lstlisting}[style=prompt]
You are an expert BigQuery SQL developer. The previous SQL
query failed with an error. Fix the query based on the
error message.

Selected Table Schema(s): {table_details}
PREVIOUS SQL (FAILED): {previous_sql}
ERROR MESSAGE: {error_message}

CRITICAL RULES:
1. Use ONLY BigQuery Standard SQL (not legacy SQL)
2. Use fully-qualified table names (project.dataset.table)
3. Map user's natural language to exact column names using
   synonyms provided
4. For derived metrics, use the formulas from metadata
5. ALWAYS include a LIMIT clause (default: LIMIT 1000)
6. Fix the specific error mentioned
7. Ensure column names exactly match db_mapping values
8. Use proper aggregation and GROUP BY clauses
9. Return ONLY the corrected SQL query, no explanations

COMMON ERRORS AND FIXES:
- "Column X not found" -> Check db_mapping, use exact name
- "Type mismatch" -> Use CAST() for date comparisons
- "JOIN error" -> Verify JOIN keys exist in both tables
- "WITH clause" -> CTEs allowed, ensure final SELECT has LIMIT

Return ONLY the SQL query.
\end{lstlisting}

\subsection{SQL G-Eval Rubric}

Used for the 100-query NL2SQL evaluation. The per-dimension breakdown (table selection 97.8\%, query understanding 99.2\%, SQL relevance 84.6\%, overall 93.9\%) is reported in Section~\ref{sec:nl2sql}.

\begin{lstlisting}[style=prompt]
You are evaluating a SQL agent's output on three dimensions.

=== INPUT ===
User Question: {user_query}
Generated SQL: {generated_sql}
Query Results: {query_results}
Final Answer: {final_answer}
Selected Table: {selected_table}
Available Tables: {available_tables}
Table Metadata: {table_metadata}

=== EVALUATION CRITERIA ===

1. QUERY UNDERSTANDING (Weight: 25%)
- Does the answer directly address the user's question?
- Is the scope correct (timeframe, geography, metrics)?
- Is all information grounded in the data?
- Is the detail level appropriate?

2. TABLE SELECTION (Weight: 35%)
Score 9-10 if correct table, 7-8 if acceptable, lower
if wrong table.

3. SQL RELEVANCE (Weight: 40%)
- Are aggregations correct (SUM, AVG, COUNT)?
- Are WHERE filters appropriate FOR THE QUESTION ASKED?
- Is GROUP BY correct?
- Do all columns exist in the table?
- Is the SQL logic sound?

IMPORTANT: Only require WHERE filters that are NECESSARY
to answer the specific question.

=== HARD-FAIL CAPS ===
- Wrong table: score <= 4
- Missing mandatory filter: score <= 5
- Non-existent column referenced: score <= 6

Return JSON with per-dimension scores (1-10), feedback,
and pass/fail flags.
\end{lstlisting}

\subsection{Insights Generation}
Used for generating the final answer to the user query post retrieval
\begin{lstlisting}[style=prompt]
You are a data insights analyst. Generate a concise summary
and highlight key trends or anomalies from the query results.

USER'S ORIGINAL QUESTION: {user_query}
SQL QUERY EXECUTED: {sql}
QUERY RESULTS (first 20 rows): {results_preview}
RESULT STATISTICS:
- Total rows: {row_count}
- Columns: {column_info}

INSTRUCTIONS:
1. Write a 2-3 sentence narrative summary answering the
user's question
2. List 2-4 key highlights as bullet points (trends,
patterns, anomalies, or notable values)
3. Focus on actionable insights
4. Mention specific numbers and metrics
5. Flag any data quality issues (nulls, zeros, unexpected
values) if present

FORMAT:
Summary: [2-3 sentence overview]
Key Highlights:
- [First key finding with specific numbers]
- [Second key finding]
- [Additional findings as relevant]
Return ONLY the formatted insights.
\end{lstlisting}

\end{document}